\documentclass[authoryear,3p]{elsarticle}
\usepackage{epsfig}
\usepackage{amsmath}
\usepackage{multirow}

\journal{New Astronomy}
\begin{document}
\begin{frontmatter}
\title{QYMSYM:  A GPU-Accelerated Hybrid Symplectic Integrator That Permits Close Encounters}
\author{Alexander Moore \& Alice C. Quillen}
\ead[url]{http://astro.pas.rochester.edu/~aquillen/}
\ead{amoore6@pas.rochester.edu \& aquillen@pas.rochester.edu}
\address{Department of Physics and Astronomy, University of Rochester, Rochester, NY 14627, USA}

\begin{abstract}
We describe a parallel hybrid symplectic integrator for planetary system integration that 
runs on a graphics processing unit (GPU).  The integrator identifies close approaches between 
particles and switches from symplectic to Hermite algorithms for particles that require 
higher resolution integrations.  The integrator is approximately as accurate as other hybrid 
symplectic integrators but is GPU accelerated.
\end{abstract}

\end{frontmatter}

\section{Introduction}
The field of solar system dynamics has a timeline of discoveries that is related to the 
computational power available (e.g., \citealt{morby01}).  As computational power has increased 
over time, so to has our ability to more accurately simulate incrementally more complex systems 
via the inclusion of more sophisticated physics or simply a greater number of particles.  The 
interaction between planets and planetesimals subsequent to formation is a well posed N-body 
problem but displays remarkable complexity even in the absence of collisions, including 
resonance capture, planetary migration, and heating and scattering of planetesimals.  In this 
paper we attempt to more accurately simulate this system in parallel by using the increased 
computational power and device memory recently made more accessible on video graphics cards.

Our code\footnote{QYMSYM is available for download at http://astro.pas.rochester.edu/\hashchar aquillen/qymsym} 
is written for Compute Unified Device Architecture (CUDA) enabled devices.  CUDA, implemented 
on NVIDIA graphics devices, is a GPGPU (General-purpose computing on graphics processing units) 
architecture that allows a programmer to use C-like programming language with extensions to code 
algorithms for execution on the graphics processing unit.  It provides a development environment 
and Application Programming Interfaces (APIs) for CUDA enabled GPUs specifically tailored for 
parallel compute purposes.  It achieves this by exposing the hardware to the developer through 
a memory management model and thread hierarchy that encourages both constant 
streaming of data as well as parallelization.  We will discuss porting the code to other parallel 
computing environments below.

\section{A second order democratic heliocentric method symplectic integrator for the GPU}
Symplectic integrators are useful for planetary system integrations because they preserve an 
energy (or Hamiltonian) that is close to the real value, setting a bound on the energy error 
during long integrations \citep{wisdom91,wisdom96}.  See \citet{yoshida90,yoshida93,leimkuhler04} 
for reviews of symplectic integrators. We have modified the second order symplectic integrators 
introduced by \citet{duncan98,chambers99}, and created an integrator that runs in parallel on a 
GPU.  We have chosen the democratic heliocentric method \citep{duncan98} because the force from 
the central body is separated from the integration of all the remaining particles and the 
coordinates do not depend on the order of the particles.

Following a canonical transformation, in heliocentric coordinates and barycentric momenta 
\citep{wisdom96} the Hamiltonian of the system can be written 
\begin{equation}
H({\bf  P, Q}) = H_{Dft}({\bf P}) + H_{Kep}({\bf P, Q}) + H_{Int}({\bf Q})
\end{equation}
where 
\begin{equation}
H_{Dft} = {1 \over 2 m_0} \left|\sum_{i=1}^n  {\bf P}_i  \right|^2 
\label{eqn:HDft}
\end{equation}
is a linear drift term and ${\bf P}_i$ is the barycentric momenta of particle $i$.
Here $m_0$ is the central particle mass.
The second term
$H_{Kep}$ is the sum of Keplerian Hamiltonians for all particles with
respect to the central body,  
\begin{equation}
H_{Kep} = \sum_{i=1}^N \left ({ {{\bf P}_i^2 \over 2 m_i} 
- {G m_i m_0  \over \left| {\bf Q}_i \right|}} \right)
\end{equation}
where ${\bf Q}_i$ are the heliocentric coordinates and are conjugate
to the barycentric momenta.  Here $m_i$ is the mass of the $i$-th particle
and $G$ is the gravitational constant.
The interaction term 
contains all gravitational interaction terms except those to the central body,
\begin{equation}
H_{Int} = \sum_{i=1}^N \sum_{j=1,j\ne i}^N -{ G m_i m_j \over 2 \left|{{\bf Q}_i - {\bf Q}_j}\right|}.
\label{eqn:Hint}
\end{equation}

A second order integrator advances with timestep $\tau$
using evolution operators (e.g., \citealt{yoshida90})
\begin{equation}
E_{Dft} \left({\tau \over 2}\right)
E_{Int} \left({\tau \over 2}\right)
E_{Kep} \left({\tau  }\right)
E_{Int} \left({\tau \over 2}\right)
E_{Dft} \left({\tau \over 2}\right)
\end{equation}
The Keplerian advance requires order $N$ computations but interaction term 
requires order $N^2$ computations.  However, encounter detection is most 
naturally done
during the Keplerian step and requires $O(N^2)$ computations if all particle pairs are
searched for close encounters.  If there is no search for close encounters then
the Keplerian and Interaction steps can be switched \citep{moore08} reducing the number of computations.

Each of the evolution operators above can be evaluated in parallel.
The drift evolution operator requires computation of the sum of the momenta.  
We have implemented this using a parallel reduction sum parallel primitive algorithm available 
with the NVIDIA CUDA Software Development Kit (SDK) 1.1 that is similar to 
the parallel prefix sum (scan) algorithm \citep{harris08}.  

The Keplerian step is
implemented by computing $f$ and $g$ functions using the universal differential 
Kepler's equation \citep{prussing93} so that bound and unbound particles can
both be integrated with the same routine (see appendix).  The Keplerian evolution step is 
also done on the GPU with each thread computing the evolution for a separate 
particle.  Kepler's equation is usually solved iteratively until a precision limit is achieved.  
However, the Laguerre algorithm \citep{conway} (also see chapter 2 by \citealt{prussing93}) 
converges more rapidly than a Newton method and moreover converges regardless of the starting 
approximation.  We have found that the routine converges to the double precision limit 
(of order $10^{-16}$) in fewer than 6 iterations independent of initial condition.  
See the appendix for the procedure.

%
%

The interaction terms are computed on the GPU with all $N^2$ force pairs 
evaluated explicitly in parallel.  The algorithm is based on the algorithm 
described by \citep{nyland08}.  This algorithm takes advantage of fast 
shared memory on board the GPU to simultaneously compute all forces 
in a $p \times p$ tile of particle positions, where $p$ is the number of 
threads chosen for the computation (typically 128 or 256).  The total energy is 
evaluated with a kernel explicitly evaluating all $N^2$ pair potential 
energy terms, similar to that calculating all $N^2$ forces.

After the change to heliocentric/barycentric coordinates, the position of 
the first coordinate corresponds to the center of mass and center of momentum.  
The trajectory of this particle does not need to be integrated.  However, it is convenient 
to calculate the energy using all pair interactions including the central mass.  
The interaction term in the Keplerian part of the Hamiltonian can be computed 
at the same times as $H_{Int}$ if ${\bf Q}_0$ is set to zero.  Consequently we set 
${ \bf Q}_0 = {\bf P}_0 = 0$ at the beginning of the computation.  This is
equivalent to working in the center of mass and momentum reference frame.
Because we would like to be able to quickly check the total energy, we have 
chosen to keep the first particle corresponding to the center of mass and 
momentum as the first element in the position and velocity arrays.  During 
computation of $H_{Int}$ we set $m_0$ to zero so that force terms from the 
first particle are not computed. These are already taken into account in the
evolution term corresponding to $H_{Kep}$.  The mass is restored during the 
energy sum computation as all potential energy terms must be calculated explicitly.

The particle positions and velocities are kept on the GPU during the computation
and transferred back into host or CPU accessible memory to output data 
files.  An additional vector of length equal to the number of particles
is allocated in global memory on the device to compute the momentum sums 
used in the drift step computation.  We limit the number of host to device and 
device to host memory transfers by persistently maintaining position and velocity 
information for all particles on the device because frequent data transfer 
between the CPU and GPU will reduce overall performance of the code.  The maximum 
theoretical throughput of PCI-express 2.0 16x bus technology, the interlink
between the CPU and GPU on Intel processor based motherboards, is 8 GB/s 
with a significant latency penalty.  This sets an upper limit of the total number of particles 
to $\sim 10^7$ based on several gigabytes of GPU memory\footnote{The linked lists 
used in the collision detection routines to be discussed enforce a lower realistic 
limit.}, but computation time on a single gpu for this number of particles 
would make such simulations unrealistic.  

While the transfer data rate limitations between the host and device are important, 
use of global device memory should also be monitored carefully.  
Depending on memory clock speed, width of the memory interface and memory 
type, theoretical maximum global memory transfer throughputs are currently 
$\sim$150 GB/s.  This is the case for the GT200(b) architecture Quadro FX 5800 and 
GeForce GTX280/285 GPUs in our cluster, which range from 140 to 160 GB/s respectively.  
Despite enjoying a significantly greater data transfer rate compared to that of the 
PCI-express interlink, the latency penalty of device memory 
is also quite large - from 400 to 800 cycles.  Therefore, explicit global memory 
access should be limited whenever possible.  Shared memory, which is basically 
a user controlled cache that can be accessed by all cores on an individual 
multiprocessor, can be used to reduce this penalty.  As described above, the 
interaction and energy kernels leverage shared memory and benefit greatly.  By 
streaming information from global memory to shared memory, we are able to hide the 
latency of global memory, further increasing the computation speed.  Even 
when all global memory transactions of a warp issued to a multiprocessor can be 
coalesced (executed simultaneously), the latency is at least the aforementioned several 
hundred cycles.  Uncoalesced global memory access can be even more costly.  
However, it is not always possible to write an algorithm to access shared memory 
in a sensible manner, and limitations on the amount of shared memory space may force 
direct global memory access regardless.  This is the case for our GPU implementation of 
the solver for the universal differential Kepler's equation.

Though the maximum number of threads on the video cards we used was 512 for GT200 
architecture cards and 1024 for GF100 architecture cards, we found that restrictions on 
the number of available registers on each multiprocessor limited all of the major kernels 
to 128 or 256 threads per block.  A more detailed review of NVIDIA GPU hardware and 
programming techniques can be found in the CUDA Programming Guide.

Our first parallel symplectic integrator necessarily ran in single precision \citep{moore08} 
as graphics cards were not until recently capable of carrying out computations in 
double precision.  Our current implementation uses double precision for all vectors 
allocated on both GPU and CPU.  A corrector has been implemented allowing 
accelerations to be computed in single precision but using double precision accuracy 
for the particle separations \citep{gaburov09}.  If we wrote a similar corrector we could 
run our interaction step in single precision (but with nearly double precision accuracy) 
achieving a potential speed up of a factor of roughly 8 on CUDA 1.3 compatible devices.  
The newest CUDA 2.0 compatible devices have superior double precision capabilities, limiting 
this potential speed up to a factor of 2.  Additionally, it would be more difficult to create 
a corrector for the Keplerian evolution step, consequently the current version of the 
integrator is exclusively in double precision.  

We work in a lengthscale in units of the outermost planet's initial semi-major axis 
and with a timescale such that $GM_* = 1$ where $M_*$ is the mass of the central star.
In these units the innermost planet's orbital period is $2 \pi$, however we often 
describe time in units of the innermost planet's initial orbital period.

\subsection{Close encounters}
Symplectic integrators cannot reduce the timestep during close encounters
without shifting the Hamiltonian integrated and destroying
the symplectic properties of the integrator (e.g., \citealt{yoshida90}). 
During a close encounter one of the interaction terms
in $H_{Int}$ becomes large compared to the Keplerian term, $H_{Kep}$.
Consequently, the symplectic integrator described above becomes inaccurate when
two massive objects undergo a close approach.
To preserve the symplectic nature of the integrator \citet{duncan98} 
used an operator splitting approach and decomposed
the potential into a set of functions with increasingly small cutoff
radii.  \citet{chambers99} instead used a transition
function which because of its relative simplicity (and reduced
numbers of computations) we have adopted here.
A transition function can be 
used to move the strong interaction terms from the interaction Hamiltonian 
to the Keplerian one so that the entire Hamiltonian is preserved 
\citep{chambers99}.
The Keplerian Hamiltonian becomes
\begin{equation}
H_{Kep}' = \sum_{i=1}^N \left ({ {{\bf P}_i^2 \over 2 m_i} 
- {G m_i m_0  \over \left| {\bf Q}_i \right|}} \right)
- \sum_{i,j=1,j\ne i}^N  { G m_i m_j \over 2 q_{ij} }  
(1 - K(q_{ij})),
\label{eqn:Hpkep}
\end{equation}
and the interaction Hamiltonian becomes
\begin{equation}
H_{Int}' = \sum_{i=1}^N \sum_{j=1,j\ne i}^N -{ G m_i m_j \over 2 q_{ij} }  
K(q_{ij})
\label{eqn:Hpint}
\end{equation}
where $q_{ij} = \left|{{\bf Q}_i - {\bf Q}_j}\right|$
and $K(q_{ij})$ is a transition or change-over function that is
zero when the distance between two objects is small and 1 when they
are large, thus $H_{Kep}' + H_{Int}'=H_{Kep} + H_{Int}$.

The transition function we use is
$$ 
K(y) = 
\begin{cases}  0                   & \mbox{if} ~~ y\le0  \\  
              \sin {y \pi \over 2} & \mbox{if} ~~ 0<y<1 \\
               1                   & \mbox{if} ~~ y\ge 1 \\
\end{cases}
\label{eqn:K}
$$ 
with 
\begin{equation}
y = {1.1 q_{ij} \over r_{crit}} - 0.1.
\end{equation}
The parameter $y$ becomes zero at $q_{ij}\sim 0.1 r_{crit}$ where $r_{crit}$ is
a critical radius.

The above transition function is
similar to that chosen by \citet{chambers99} but
we use a sine function instead of a polynomial function
as sine is computed quickly on graphics card (estimated
at 4 or 5 floating point computations rather than the 8
required for the polynomial function used by \citealt{chambers99}).

\subsection{Hermite integrator}
The numerical integrator used for particles undergoing close approaches also
must be reasonably fast as we would like to make it possible
for our integrator to integrate as many particles as possible.
Consequently we have chosen a 4-th order adaptive step size Hermite
integrator for particles undergoing
close approaches instead of the Burlisch-Stoer integrator used
by \citet{chambers99}.  This integrator forms the heart of many
N-body integrators (e.g., \citealt{makino03}).
Our code follows the algorithm described by \citet{makino92}
but does not use the Ahmad-Cohen scheme.    While the Hermite integrator runs
on the CPU, we have implemented the routine to integrate the accelerations and
jerks both on the CPU and GPU.   The GPU version is called if the number of particles 
integrated exceeds a certain value, $np_{switch}$.

The Hermite integrator must be modified to use 
the transition function for computation of
both accelerations and jerks (time derivatives of the accelerations).
Using predicted velocities we evaluate the acceleration $a_i$
of particle $i$ according to the following
\begin{eqnarray}
{\bf a}_i &=& 
\sum_{j>0} 
G m_j 
{{\bf q}_{ij} \over  s_{ij}^{3}} 
\left[1-K(s_{ij}) + s_{ij} K'(s_{ij}) \right] \nonumber \\
&& ~~~~ + 
G m_0 {{\bf q}_{i0} \over s_{i0}^{3}} 
\end{eqnarray}
where ${\bf q}_{ij} = {\bf q}_i - {\bf q}_j$.   Here 
$s_{ij} = \sqrt{q_{ij}^2 + \epsilon^2}$ and $\epsilon$ is an optional smoothing length  
(see equation 2 by \citealt{makino92} for comparison).  
The jerks are evaluated with 
\begin{eqnarray}
\dot{\bf a}_i &=& 
\sum_j {G m_j \over s_{ij}^3} \left\{
({\bf v}_{ij} \cdot {\bf q}_{ij}) {\bf q}_{ij} K''(s_{ij}) + \right. \nonumber \\
&&  
\left. \left(
   {\bf v}_{ij}   -{3 ({\bf v}_{ij} \cdot {\bf q}_{ij}) {\bf q}_{ij} \over s_{ij}^2}  
\right) 
     \left[1- K(s_{ij}) + s_{ij} K'(s_{ij})\right]\right\}  \nonumber \\
 & &           +{ G m_0 \over s_{i0}} 
 \left(  {\bf v}_{i0}   -{3 ({\bf v}_{i0} \cdot {\bf q}_{i0}) {\bf q}_{i0} \over s_{i0}^2} \right) 
\end{eqnarray}
where ${\bf v}_{ij} = \dot {\bf q}_i - \dot {\bf q}_j$ are the difference between predict
position and velocities vectors, respectively, for particles $i$ and $j$;

The transition radius 
$r_{crit}$ was described by \citet{chambers99} in terms of the mutual Hill radius or 
\begin{equation}
R_{MH} \equiv \left({m_i + m_j \over 3 M_0}\right)^{1/3} 
\left[{a_i + a_j \over 2}\right]
\end{equation}
though \citet{chambers99} also included a expression that depends on the speed.
One problem with this choice is that the force between particles don't solely depend on the distance 
between the particles as they also depend on the distance to the central star.  This implies that the 
forces are not conservative and makes it more difficult to check the energy conservation of the Hermite 
integrator.  Instead we choose a critical radius, $r_{crit}$, prior to the Hermite integration and keep 
it fixed during the integration.  The critical radius $r_{crit}$ is defined to be a factor $a_H$ times 
the maximum Hill radius of the particles involved in an encounter at the beginning of the Hermite integration.  
\begin{equation}
r_{crit} = a_H \times \max 
\left\{ r_{H,i} ~~ \mbox{for} ~~  i  ~~ \mbox{in ~ encounter~ list}
\right\}
\label{eqn:rcrit}
\end{equation}
where $r_{H,i}$ is the Hill radius of particle $i$.

When using the Hermite integrator we do not allow the central star to move as we 
keep the system in heliocentric coordinates.  We have checked that the total energy 
given by $H'_{Kep}$ (equation \ref{eqn:Hpkep}) is well conserved by the Hermite integrator.  
We find that the accuracy is as good as that using the Hermite integrator
lacking the transition function and is set by the two parameters controlling
the timestep choice $\eta$ and $\eta_s$ (see section 2.1 and equations 7, and 9 \citealt{makino92}).

We note that our modified interaction step requires $N^2$ computations and adding in the transition function 
would substantially add to the number of computations involved in computing accelerations.  We instead make 
use of the list of particles identified during our encounter identification routine
to correct the forces on the particles that are involved in encounters.   Consequently all interactions are 
computed and then only those involved in encounters are corrected just before the Hermite integrator is called.

\subsection{Integration procedure for the Hybrid integrator}

Our procedure for each time step is as follows:
\begin{enumerate}
\item
 Do a drift step (evolving using $H_{Dft}$; equation \ref{eqn:HDft}) for all particles for timestep $\tau/2$ on GPU.
 \item
 Do an interaction step (evolving using $H_{Int}$; equation \ref{eqn:Hint}) using all particles and for all interactions
 and without using the transition function $K$ for timestep $\tau/2$ on GPU.
  \item
 Do a Keplerian step (evolving using $f$ and $g$ functions) for all particles for timestep $\tau$ on the GPU. Note that the ones undergoing close approaches have been inaccurately integrated but will be corrected later.     
\item
 Use stored positions and velocities 
(${\bf q}_0, {\bf v}_0, {\bf q}_1, {\bf v}_1$) 
prior and after the Keplerian step in global memory on the GPU 
to identify close encounters on the GPU.  
If there are encounters, transfer pair lists onto the CPU and divide 
the list of particles involved in encounters into 
non-intersecting sets.  Calculate the maximum Hill radius for particles in each encounter list and use this radius to compute a critical radius $r_{crit}$ for each encounter set. 
Only if there are encounters
 are the positions and velocities prior and after the Kepstep copied 
onto the CPU.  
  \item 
  For each encounter set, subtract interactions that should not have been 
previously calculated in step \#2 using stored positions and velocities 
on the CPU.  Previously we calculated the interaction step using all interactions.   
However we should have weighted them with a transition function $K$ (equation \ref{eqn:K}).  Now that we have a critical radius, $r_{crit}$, estimated for each encounter list we can correct the interaction step. 
After interactions weighted by $1-K$ have been subtracted, the interaction step effectively calculated is  $H'_{Int}$ (equation \ref{eqn:Hpint}) rather than $H_{Int}$ (equation \ref{eqn:Hint}).   
\item
 Use a modified Hermite integrator 
to integrate close approach sublists for $\tau$ using
the stored CPU positions and velocities ${\bf q}_0, {\bf v}_0$.
\item 
For each encounter set, subtract interactions that will be incorrectly calculated by a repeat of the interaction step in step \#9.
\item
If there have been encounters,
copy only particles involved in encounters back into the arrays ${\bf q}_1$ and ${\bf v}_1$ on the CPU.  Copy the entire arrays on the the GPU.   Particles involved in encounters have been integrated using a Hermite integrator.  Particles not involved in encounters have had Keplerian evolution only.
 \item
 Do an interaction step (using $H_{Int}$) using all particles and for all interactions
 and without using the transition function $K$ for timestep $\tau/2$.
 \item
 Do a drift step (using $H_{Dft}$) for all particles for timestep $\tau/2$.
 \end{enumerate}
 
As we have checked for encounters during every time step we can flag  encounters involving more than one planet.  In this case we can choose to do the entire integration step for all particles with the Hermite integrator (but with accelerations and jerks
computed on the GPU).  As planet/planet encounters are rare this does not need
to be done often.   Hybrid symplectic integrators such as {\it Mercury} \citep{chambers99} 
have largest deviations in energy during infrequent planet/planet encounters.   By integrating the entire system with a conventional N-body integrator during planet/planet encounters we can improve the accuracy of the integrator without 
compromising the long term stability of the symplectic integrator.

\subsection{Identifying close encounters}

Close encounter identification is in general an order $N^2$ computation
as all particle pairs must be checked every timestep.
This is potentially even more computationally intensive than
computing all interactions.
We do this with two sweeps, each one considering
fewer particle pairs.
The first sweep is crude, covers all possible particle pairs and so
is order $N^2$. This one should be as fast as possible to 
minimize its computational intensity.  Shared memory is
used for particle positions 
in a tile computation similar to that used to compute
all force interactions.  The computation 
can be done with single precision floating point computations and we
can be conservative rather than accurate with encounter identification. 
For each particle we compute 
its escape velocity (and this is order $N$ as it is only
done for each particle).  A particle pair
is counted if
the distance between the particles is smaller than the sum of a factor times the mutual Hill radius times the distance moved by the first particle
moving at its escape velocity during the timestep. 

The first kernel call sweeps through all particle pairs but only counts the
number of possible interactors.   An array of counts 
(one per particle) is then scanned
in parallel using the parallel prefix (scan) function \citep{harris08} available
in the CUDPP subroutine library.
CUDPP is the CUDA Data Parallel Primitives Library and is 
a library of data-parallel algorithm primitives such as 
parallel prefix-sum, parallel sort and parallel 
reduction.\footnote{http://gpgpu.org/developer/cudpp}
The second kernel call then uses
the scanned array to address locations to record 
pair identification numbers identified in the crude sweep.
As the number of pairs is identified during the first kernel
call memory requirements can be considered before calling
the second kernel call.

A second more rigorous sweep is done on the pairs identified from
the first one.
For this sweep we use the particle positions and velocities
computed using Keplerian evolution at the beginning
and end of the timestep.
We use the third order interpolation scheme described in section 
4.4 by \citet{chambers99} to 
predict the minimum separation during the timestep
which we compare to the sum of their Hill radii.
Pairs which fail this test are marked.   
Pairs which approach within a factor $a_E$ times times the sum
of their Hill radii are marked as undergoing a close encounter.
Using a second
scan we repack the list of identified pairs into a smaller array.

After all pairs undergoing encounters  have been identified,
we sort them into non-intersecting sublists.  This is done
on the CPU as we expect the number of pairs now identified is not
large.

We note that since we used the escape velocity in our
first sweep to identify pairs
of particles undergoing encounters, we could miss encounters between
particles escaping from the system and other particles.
The number missed we suspect would be small.

The number of pairs identified in the first crude sweep depends
on the timestep and the particle density.   We could consider
other algorithms for removing possible particle pairs
from consideration as long as we keep in mind that
the first sweep should remove as many pairs as possible
while being as efficient as possible.

\subsection{List of Parameters}
We review some parameter definitions
\begin{enumerate}
\item 
The timestep $\tau$.   As the symplectic integrator is second
order the error should depend on $\tau^3$.
\item
The Hill factor 
$a_H$ is used to define $r_{crit}$ (equation \ref{eqn:rcrit}). This parameter 
is needed to compute the transition function $K$ (equation \ref{eqn:K})
and so is needed by the Hermite integrator and to compute interactions
when there are close encounters (equation \ref{eqn:Hpint}).
This parameter is a distance in Hill radii of the most massive particle
involved in a close encounter.
\item
The Hill factor $a_E$.  Particle pairs with minimum estimated approach distances
within $a_E$ times the the sum of their Hill radii are identified
as undergoing close approaches.
\item
The smoothing length, $\epsilon_H$, used in the Hermite integrator.
As we do not yet take into account actual collisions, this parameter 
should be small but not zero.
A non-zero smoothing length will prevent extremely 
small timesteps in the event of a close approach of two point masses. 
\item The parameters setting the accuracy of the Hermite integrator
$\eta$, and $\eta_s$ (as discussed and defined by \citealt{makino92}).
\item The number $np_{switch}$. If the number of particles 
involved in an encounter is larger than this number 
then the Hermite integration
is done on the GPU rather than on the CPU.
\end{enumerate}

\section{Test Integrations}
The chaotic nature of the many-body problem makes it challenging to check 
the accuracy of any code designed to simulate solar or extrasolar systems.
There is no simple way to generate analytical solutions to a set of initial 
conditions.  However, it is possible for us to run our integrator through a suite of tests and 
compare those results to the integrators that form the basis or our code, 
namely the the hybrid symplectic integrator {\it Mercury} by \citet{chambers99} 
and the democratic heliocentric modification to mixed variable symplectic 
integrators \citet{duncan98} implemented in the {\it SyMBA} package.

One of the most basic tests we ran on the integrator was to check that 
smaller timesteps ensured greater energy conservation for a given set of 
initial conditions.  We use the relative energy error as a metric for 
measuring the conservation of energy in a simulation.  This relative 
energy error is computed with the formula $\Delta E = (E - E_0) / E_0 $
where $E_0$ is the energy at the beginning of the computation.  
Indeed, we do observe superior energy conservation with smaller timestep 
sizes with the error scaling $O(\tau^3$) as expected as the integrator
is second order.  Two simple test integrations of 1024 particles and 
identical initial conditions with timesteps of 0.1 and 0.01 were completed.  
Over 10 timesteps of 0.1 the relative energy error 
was $\Delta E = -8.429 \times 10^{-6} $ with a per step average energy error of 
$\Delta E = -8.429 \times 10^{-7}$.  Over 100 timesteps of 0.01 the relative energy error 
was $\Delta E = -8.422 \times 10^{-8} $ with a per step average energy error of 
$\Delta E = -8.422 \times 10^{-10}$.  Comparing the average energy error per step for 
these two test simulations indicates an exact scaling with $\tau^3$.

\subsection{Enhanced or Scaled Outer Solar System}
\begin{figure}
\includegraphics[scale=1.0]{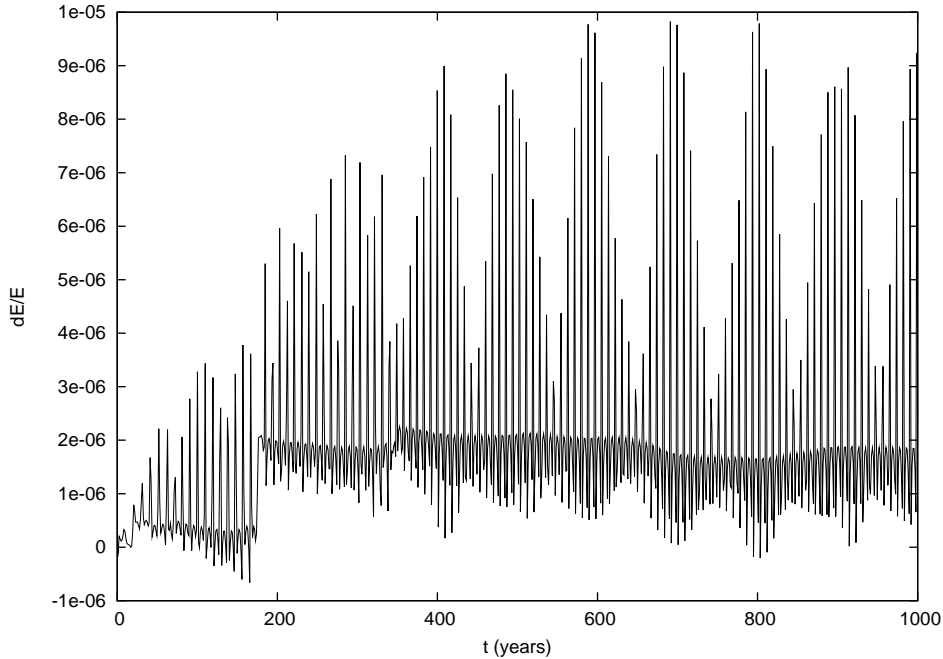}
\caption{
We show the relative energy error $(E-E_0)/E_0$ from an integration 
of the outer 
solar system (Jupiter, Saturn, Uranus, Neptune) only with masses enhanced by 
a factor of 50.  
Time is in units of the orbital period in years of the innermost planet.
The energy errors are bounded.  This behavior is typical of 
fixed timestep symplectic integrators.
There is a planet/planet encounter at a time of about 200 years.
Eventually planets are ejected.
\label{fig:enhanced_outer_solar_system}
}
\end{figure}

Our primary test integration is an integration of a scaled 
version of the outer solar system.  Both $SyMBA$ and $Mercury$ were tested in 
this manner \citep{duncan98,chambers99}.  The simulation consists of the four 
giant planets in the Solar system but with masses 
increased by a factor of 50.  Previous simulations demonstrate that this 
configuration is unstable \citep{duncan97}, although they disagree on the eventual outcome 
due to the chaotic nature of solar system evolution.  As noted by 
\citet{chambers99}, the timestep chosen can have an effect on 
the eventual outcome even when varied only slightly.  With these facts in 
mind, we ran a simulation with values as close as possible to 
those of previous tests.  To match the simulation described
in section 5.1 by \citet{chambers99} we used a timestep of 
$\tau = 0.00255 P_0$ where $P_0$ is the initial orbital period of the innermost
planet.  We set the Hill factor $a_E = 2.5$ for encounter detection
and $a_H = 1.0$ setting the transition radius, $r_{crit}$.
For initial conditions we used epoch J2000.0 orbital 
elements for the four giant planets \citep{standish92}.
The energy error for this simulation is shown in Figure \ref{fig:enhanced_outer_solar_system}.  
Time is given in orbital periods of the innermost planet.
We ran our simulation for the same number of orbital periods
as did \citet{chambers99} in the test shown in their Figure 2.
 
Despite the highly chaotic and unstable nature of this system, our integration 
is remarkably quite similar to that by 
\citet{chambers99} (see their Figure 2). 
The energy error is bounded, as expected for a symplectic integrator.
The spikes in the energy error are also evident with other integrators
(see Figure 2 by \citealt{chambers99} and Figure 6 by \citealt{duncan98}).
A close encounter between Jupiter and Saturn is experienced at approximately 
200 years into the integration, which causes a jump in the relative energy 
error.  A similar jump in energy was also seen by \citet{chambers99}.  
Our simulation also involved the later ejection of more than one planet.

The energy error of our integration is bounded - typical of a 
fixed timestep symplectic integrator.  However the sizes of the
individual spikes in energy error is larger than those shown 
in Figure 2 by \citet{chambers99} though it is similar in size
to those shown in Figure 6 by \citet{duncan98}.
The spikes in Figure 2 by \citet{chambers99} are of order $10^{-6}$ 
in fraction error whereas ours are of order $10^{-5}$.
There are several possible explanations for the worse performance of our
integration.  During close approaches we are using a Hermite integrator
rather than the Burlisch-Stoer used by \citet{chambers99}.  
However we have measured the error across 
each close approach and find conservation of $H_{Int}'$ at a level
orders of magnitude below $10^{-5}$ so the choice of integrator for
close approaches is unlikely to be the cause.  We have checked our drift and Keplerian
evolution operators and find they conserve their Hamiltonians within
the precision of double precision arithmetic.  We find
little dependence on energy error in the form of the transition function,
or the Hill factor $a_H$ setting $r_{crit}$. 
However the Hill factor influencing the identification 
of close encounters, $a_E$, does affect the energy error.   
During encounters, terms in the interaction Hamiltonian 
(equation \ref{eqn:Hint}) can become large.   However we add these
into the interaction term during the interaction evolution.
These terms are then removed subsequently once encounters are identified so
that $H_{int}'$ is calculated (see the discussion in section 
2.1 and steps 3 and 7 in section 2.3). This 
procedure for removing the incorrectly calculated
interaction terms could account for the somewhat
poorer performance of our integrator. 

\subsection{Long term outer Solar system evolution}
For our second integration we compare the relative energy error of the evolution of the 
outer Solar system with much larger timesteps and for a much larger amount of time - 
similar to the test integration discussed by \citet{duncan98} and shown in their Figure 2.  
In this test, we again use the epoch J2000.0 orbital elements as initial conditions but use 
a timestep of $\tau = 0.0318 P_0$ where $P_0$ is the initial orbital period of the innermost 
planet.  The timestep used and the total integration time ($\sim 3 \times 10^5$ yr) are similar to 
those values used in \citet{duncan98}.  The masses of the planets are unchanged from their 
accepted values.  This configuration is known to 
be stable so we did not expect any close encounters or ejections and a somewhat 
better relative energy error compared to the last simulation despite the increase 
in timestep size.  As in the previous test, we did not observe a relative energy 
error as good as \citet{duncan98}; our accuracy being a factor of a few poorer 
in terms of both the average energy error as well as the size of the fluctuations 
in the energy error. No encounters are present in this simulation suggesting
that we have somewhat larger sources of errors
in our interaction or Keplerian evolution steps than {\it SyMBA}. 
We are not yet sure what is causing this low level of error 
as these computations have been done in double precision and when 
tested individually for single timesteps, we have found them 
accurate.

\begin{figure}
\includegraphics[scale=1.0]{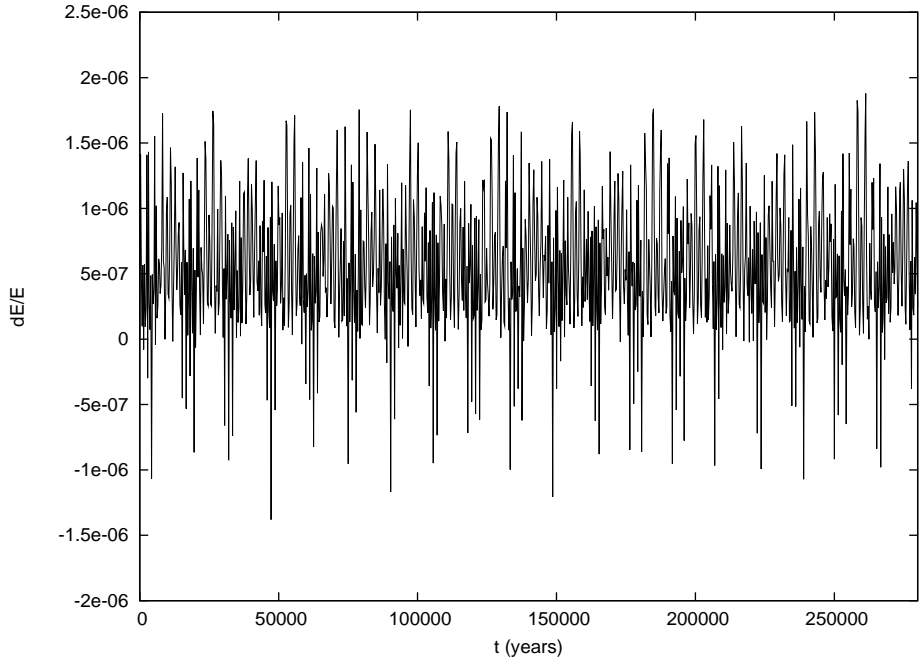}
\caption
{
We show the relative energy error in a simulation of the outer 
solar system (Jupiter, Saturn, Uranus, Neptune).  Initial orbital elements are 
those of the giant planets
at epoch J2000.0.  Time is given in units
of the orbital period in years of the innermost planet.
\label{fig:outer_solar_system}
}
\end{figure}

\subsection{Sensitivity of Parameters}
We find that the energy error is insensitive to the Hill factor $a_H$ setting
the transition function but is quite sensitive to $a_E$, the Hill factor
for encounter detection.  We find the best numerical results for $a_E$
in the range 1 to 4.  The larger the value of $a_E$ the more
encounters are sent to the Hermite integrator and the slower the integration.
However if $a_E$ is too small then the difference between the integrated
Hamiltonian and true one will be large as the interaction terms become large.

\section{Benchmarks and Profiling}
In this section we discuss the fraction of runtime spent doing each computation in 
some sample integrations.  Computations that are run on the GPU are called kernels.  
In Tables \ref{tab:tab1} and \ref{tab:tab2} we list the fraction of GPU time spent 
in each kernel or group of kernels or doing memory operations for these two different 
simulations.  We also list the CPU runtime for each kernel which includes 
the overhead for calling the device function in addition to the runtime on the GPU.  
We label the kernels or groups of kernels as follows: 
Interaction (evolving using $H_{int}$),
Keplerian (evolving using $H_{Kep}$),
Sweeps (the kernels for finding close encounters),
Drift (evolving using $H_{Dft}$),
Energy (evaluating the total energy on the GPU and only done once per data output), and
Hermite (when the Hermite integrator is run on the GPU).
Also listed is the total time spent doing
memory allocation and transfers (listed under `Memory').
All other kernels are listed under `Various' and the computation times
are summed.  Kernels in the `Various' category include center of momentum calculations, 
scans, repacking, etc that are not part of the previous listed kernels.  
Three sweeps are called, the first two passing over
all particle pairs.  The sum of the time spent in all sweeps is shown
in the table under Sweeps.

The simulations described in Table 1 and 2 are identical except in the number of particles in the simulation.  
The initial conditions consist of 4 massive planets inside a debris disk which is significantly less massive than 
the planets and is truncated relatively quickly.  Both simulations were run for 1 output of 100 timesteps.  
We note that the energy kernel is only called once per output.
The set of initial conditions was chosen so that the system would quickly have close encounters involving two planets and 
so force the integrator to call the Hermite integrator with all particles.  This allowed us to measure the performance 
of different kernels.  Relevant profiler information includes the number of kernel calls made of a particular 
function, total time to complete all calls of that function on both the CPU and GPU and the percentage of GPU 
runtime spent running each kernels.  Tables 1 and 2 represent a scenario in which collision are frequent and 
the Hermite integrator is called frequently.

Profiling for the code was done with the NVIDIA CUDA Visual profiler version 3.0 that is available with 
the CUDA toolkit.  Profiling was done on two video cards, the NVIDIA GeForce GTX 285 and the GeForce GTX 480.  
Tables 1-2 only show the times for computations on the GPU alone and CPU overhead + GPU time to call these 
functions, but do not show the fraction of total computation time on the GPU.  However, using the GPU Utilization 
Plot available in the Visual profiler, we are able to determine the session level (an entire integration) GPU 
utilization.  For $10^3$ particles we achieved a utilization of 93\% and for $10^4$ particles a utilization of 
85\%.  For $10^5$ particles, utilization varied dramatically, but never dipped below $\sim$50\%.  The overall 
utilization dropping for higher number of particles may seem counterintuitive, but makes sense in light of 
the increasing number of particles that are flagged for close approaches and sent to the CPU for integration.  
In fact, in certain cases, particularly with low densities or small hill factor identification radii, $10^5$ 
particle simulations would display very high utilization.  It is important to note that these utilization 
percentages represent the time that the GPU is not idle - it is not indicative of the actual performance of a 
particular kernel or the code as a whole.  There are other more useful metrics for individual kernel performance.
Also, we note that the Hermite integrator is the only major routine in the code that is run on the CPU and 
because of this the host processor speed and quantity of main system memory have little effect on the values 
given in the tables below.  The effect of the higher CPU speeds is to decrease the total runtime and to effect 
the GPU utilization values.  Only in simulations with lower GPU utilization percentages do the effects of 
increased processor speed become apparent.  Altering main system memory amounts or speeds has no practical 
effect on our runtime due to the relatively small amount of memory being used - even for $10^5$ particle simulations.

We observed that nearly all of the runtime is dominated by the interaction step, the Hermite integrations and the 
close approach detection kernels labeled "Sweeps" in the table.  Memory operations, the drift and Keplerian 
evolution steps of the symplectic integrator and all of the other various functions on the GPU add up to only a 
small fraction of the runtime.  This percentage will continue to decrease as the particle number is increased as 
the interaction step is $O(N^2)$ but the Keplerian evolution and drift step are $O(N)$.  The ratio of time spent 
doing sweeps compared to that in the interaction step may increase with $N$ as there may be more encounters 
when the particle density is higher.  Comparing Table \ref{tab:tab1} and \ref{tab:tab2} we see that the fraction
of time spent in the interaction step is only somewhat larger when the number of particles is larger.  This implies 
that the fraction of time spent doing operations that are $O(N)$ is small even when the number of particles is only 1024.

Examining the timing information from the simulations with 10240 particles we note that on average a single 
call to the Hermite and interaction kernels took about 0.20 and 0.11 seconds respectively.  We also observe 
that the CPU + GPU runtime are nearly identical to the GPU runtime alone for all major kernels.  It is only 
with smaller numbers of particles or kernels that are $O(N)$ that these values diverge even slightly.

In Table \ref{tab:tab3} we show a comparison of simulations with three different particle numbers and 
identical initial conditions as those given for the first two sets of simulations described in 
the first two tables.  However, these were evolved for only 10 timesteps instead of 100 before outputting 
data.  These integrations lack close encounters and so serve to compare sweep, energy and interaction kernels.  
For simulations with sparse debris disks or low mass objects, 
this third table describes how the GPU runtime will be spent.  Clearly, the interaction step, energy calculation 
and sweeps compose a majority of the runtime.  As larger number of timesteps are taken per data out or if we 
suppress the calculation of the energy, we will reach an asymptotic limit of ~85\% GPU time spent on interaction 
step with the remaining ~15\% of the GPU time spent on the sweeps for encounter detection.\footnote{GPU time in 
$\%$ listed by the CUDA profiler does not always add up to 100 for reasons of significant figures.}

\begin{table}
\caption{\large Profile for 1024 particles \label{tab:tab1}}
\begin{tabular}{@{}llccc}
\hline
Function/Kernel & No. of calls	& GPU Time($\mu$s)	& CPU Time($\mu$s)	& GPU time (\%) 	\\ 
\hline
Interaction 	& 200 		& 790202		& 792607		& 51.40			\\
Hermite 	& 80 		& 537281		& 539429		& 34.95			\\
Sweeps 		& 300 		& 147786		& 151312		& 9.59			\\
Mem. Trans.	& 2566 		& 22567			& 41226			& 1.46			\\
Keplerian 	& 100 		& 17116			& 18308			& 1.11			\\
Energy 		& 3 		& 11227			& 11287			& 0.73			\\
Drift 		& 200 		& 1770			& 4007			& 0.11			\\
Various 	& 1521 		& 9217			& 27259			& 0.57			\\
\hline
\end{tabular}
{\\ This table shows the fraction of time spent in
different tasks  for an integration of 1024 particles integrated for  
100 timesteps on an NVIDIA GTX 285.   The leftmost column lists
the Kernels.  `Mem. Trans.' denotes the time involved in memory transfers.
}
\end{table}

\begin{table}
\caption{\large Profile for 10240 particles \label{tab:tab2}}
\begin{tabular}{@{}llccc}
\hline
Function/Kernel & No. of calls	& GPU Time($\mu$s)	& CPU Time($\mu$s)	& GPU time (\%) 	\\ 
\hline
Interaction 	& 200 		& $2.29\times 10^7$	& $2.29\times 10^7$	& 53.21			\\
Hermite 	& 80 		& $1.60\times 10^7$	& $1.60\times 10^7$	& 37.20			\\
Sweeps 		& 300 		& $3.58\times 10^6$	& $3.59\times 10^6$	& 8.30			\\
Energy 		& 3 		& 302848		& 302920		& 0.70			\\
Mem. Trans.	& 2566 		& 171014		& 316417		& 0.39			\\
Keplerian 	& 100 		& 48186			& 49397			& 0.11			\\
Drift 		& 200 		& 8183			& 10427			& 0.01			\\
Various 	& 1721 		& 15529			& 35350			& 0.02			\\
\hline
\end{tabular}
{\\ Similar to Table \ref{tab:tab1}.  
This profile is for 10240 particles integrated for 100 timesteps on an NVIDIA GTX 285.}
\end{table}

\begin{table*}
\caption{\large Comparing Kernel speeds for $10^3$, $10^4$ and $10^5$ particles\label{tab:tab3}}
\begin{tabular}{@{}lllll}
\hline
Kernel  	& No. Particles 	& No. of calls	& Time $\mu$s	& GPU time (\%) 	\\ 
\hline
\hline
\multirow{3}{*}
{Interaction}	& 102400		& \multirow{3}{*}{20}	& $1.965\times 10^8$	& 77.26			\\
            	& 10240			&    		& $2.286\times 10^6$	& 76.91			\\
            	& 1024			&    		& 79019		& 71.84			\\
\hline
\multirow{3}{*}
{Sweeps}	& 102400		& \multirow{3}{*}{30}	& $3.071\times 10^7$	& 12.06			\\
       		& 10240			&    		& 356295	& 11.98			\\
       		& 1024			&    		& 14776		& 13.41			\\
\hline
\multirow{3}{*}
{Energy}	& 102400		& \multirow{3}{*}{3}	& $2.689\times 10^7$	& 10.59			\\
       		& 10240			&   		& 307708	& 10.35			\\
       		& 1024			&   		& 11210		& 10.19			\\
\hline
\multirow{3}{*}
{Various}	& 102400		& 507		& 234894	& 0.08 			\\
        	& 10240			& 477		& 21849		& 0.70			\\
        	& 1024			& 457		& 4986		& 4.51			\\
\hline
\end{tabular}
{ \\
This profile shows something akin to the asymptotic limit of simulations which do not have
objects experiencing close approaches.  Each simulation is of a single 
output of 10 timesteps on a 285 GTX with the specified number of particles.}
\end{table*}

\subsection{Optimizations}

We have optimized our code beyond the standard CPU software optimization techniques 
by using some of the ``best practices'' for GPU programming.  Refer to the "Best Practices 
Guide -- CUDA 3.0" for an in depth description of low, medium and high priority optimizations.  
The most important and most obvious best practice is to run as much of your code on the 
GPU as possible while simultaneously implementing each kernel to take advantage of as much 
parallelism as possible.  To this end, we have put nearly all of the computations for our 
integrator on the GPU including all the evolution operators in our symplectic integrator 
(drift, Keplerian, interaction), all collision detection sweeps, the Hermite integration 
routine as well as the energy computation.  As shown in the profiling, the execution time 
of a non-interacting system is dominated by the $O(N^2)$ interaction term; 
a kernel with a high degree of arithmetic intensity and minimal memory transfer compared 
to the number of floating point operations.  Some of these routines show small GPU 
performance benefits over the CPU because of their large size in terms of registers, lack of arithmetic 
intensity in that they are $O(N)$, and their inability to use shared memory.  However, 
even these serial routines are executed on the GPU to ensure as few host to GPU or GPU to 
host memory transfers.  This best practice is very important because it ensures fewer high 
latency memory transfers between the system memory and global memory on the GPU and again 
from global memory to the multiprocessors.  The more time that can be spent doing floating 
point operations rather than memory transfer allows for the greatest GPU speed increases.  
Other high priority optimizations are to access shared memory over global memory whenever 
possible and to keep your kernels from having diverging execution paths.  Some routines 
were not or could not be written to use shared memory, but all $O(N^2)$ sweeps and the 
interaction step leverage this tremendous performance enhancer.  All kernels are written 
to minimize branching statements to ensure that execution paths do not diverge.  
This best practice manifests itself most obviously in the interaction detection and Hermite 
integration kernels.  Collision detection was not incorporated in the same function, but 
rather in a separate kernel.  In another example of code design choice, we over count and 
do $N^2$ computations in the interaction step instead of $N (N-1)/2$ since it allows for 
us to have simpler code that is more easily executed in parallel on the GPU.

Medium priority optimizations including use of the fast math library that could potentially 
effect the accuracy of our code in a negative way are not implemented.  We do use multiples 
of 32 threads for each block and we are able to attain a 33\% occupancy rate for the 
interaction step, the energy computation and the GPU version of the Hermite gravity step as 
well as a 50\% occupancy rate for 
the first and second sweep steps on a GF100 based GTX 480.  The concept of occupancy is a 
complicated one and is explained in more detail in both the "Best Practices Guide -- CUDA 3.0" 
guide and the "CUDA Programming Guide 3.0".  At it's core it's simply a ratio of the number 
of active warps in a multiprocessor to the maximum number of possible warps the multiprocessor 
can maintain.  A warp is simply a group of 32 threads to be executed on a multiprocessor.  
While computing the number of active warps per multiprocessor involves hardware 
knowledge beyond this document, and noting that higher occupancy does not necessarily equate 
to better performance, it is important to maintain a minimum occupancy.  Below some value, the 
latencies involved with launching warps can not be hidden.  The value suggested in the 
"Best Practices Guide -- CUDA 3.0" suggest an occupancy of at least 25\% be maintained.  
As mentioned, we are able to attain this.

We also attempt to ensure optimal usage of register space per block and to loop unroll 
functions when it can increase performance.  The number of loop unrolls that can be achieved is 
determined by examining the numbers of registers used per thread for a given 
function and multiplying it by the number of threads per block that you wish to issue.  
This gives the total number of registers used per multiprocessor - the group of CUDA 
cores that a block is issued to.  This value must be smaller than the number of registers 
available on the relevant architectures multiprocessor.  Increasing the number of 
loop unrolls in a given kernel can alter the register requirements, so there is a limit to the 
number of loop unrolls that can be implemented.  For reference, the GT200 architecture (GTX 285) and the 
GF100 architecture (GTX 480), there are 8 and 32 CUDA cores per multiprocessor, 16K
and 32K 32-bit registers, and 16KB and 48KB per multiprocessor of shared memory 
respectively.\footnote{Double precision values are 64-bit, requiring 2 registers per value.}
All relevant hardware numbers including number of threads, multiprocessors, cores per 
multiprocessor, amount of shared memory, etc. can be found in the CUDA Programming Guide 
in appendices A and G.
It is possible to artificially restrict a function to use a number of registers 
that is less than it requires.  However, this forces some values to be stored in 
local memory (global memory), which, as mentioned previously, carries a very heavy performance 
penalty in terms of latency.  For this reason, it can be detrimental to limit the 
number of registers per thread below that which is required.  We found that our performance 
was best with a loop unroll of 4, a maximum number of registers per thread set to 64, 
and 128 threads per block.

\subsection{Use of Parallel Primitives}

The parallel computations used by this code for the most part utilize
parallel primitives and so can be ported to other parallel computation platforms
with similar parallel primitive libraries.   The energy computation, 
interaction step and all pairs sweep identification routines essentially 
utilize the same tiled shared memory algorithm (described by \citealt{nyland08}.
The drift step and encounter detection use a parallel prefix sum 
(see \citealt{harris08}).

\subsection{Areas for future optimization}

There are several areas in which our code could be further optimized.  Without drastically 
altering the code there are some "low priority" optimizations such as the use of 
constant memory for unchanging values like smoothing lengths and the Hill factors.

More important hardware level optimizations could include alternate ways to ensure global memory coalescing 
and prevent shared memory bank conflicts by padding our current data structures or disassembling them 
entirely and using single arrays of double precision elements to store data. 

On a higher level, there are several optimizations that could possible speed up the code 
significantly.  First, replacing the interaction step which is currently an 
all-pairs $O(N^2)$ calculation with a GPU enhanced tree integration 
(e.g., \citealt{richardson00,gaburov10}) could bring a speed increase to the code 
and allow larger numbers of particles to be simulated.

Additionally, the sorting routines ('sweep kernels') could be optimized to use faster detection 
methods than the current $O(N^2)$ methods.  By implementing a parallel sorting algorithm or 
simply making our detection routine more intelligent, we could reduce this part of the runtime.  
For simulation of many massless but colliding particles (such as dust particles) in the vicinity 
of planets and planetesimals an improvement in the encounter identification may allow us to integrate
many more particles.

A potentially simpler change would be the implementation of double precision in software rather than 
hardware.  Double precision code on NVIDIA GPU's executes more slowly than single precision, as is 
often the case in hardware.  Depending on the device, this factor can be anywhere 
from 1/8th to 1/2 the speed and some older CUDA capable devices do not support double 
precision at all.  By implementing double precision in software we would be able to 
compile our code using 32-bit floating point precision.  This would have the benefit of 
multiplying the speed of the code you are running by a factor of at least a few and 
could possibly allow devices not originally capable of supporting the code to run.  
Additionally, it appears that only GF100 devices in the Tesla brand of cards will 
support full double precision speeds which run at 1/2 the number of double precision flops.  
A penalty of ~1/8th the number of single precision flops was observed while profiling our 
code on a GTX 480.  This was observed indirectly by only noticing a speed increase of ~2x 
on our GF100 based machine for our double precision kernels like the interaction step and 
the energy computation over our GT200 based GPUs.  At 1/8th the peak double precision flops, 
the GF100 based GPU should be ~2x faster than a GT200 based GPU because it has 2x the 
number of cores over the GT200 based cards (ignoring minor execution time differences 
based on the actual core clock speeds).  The GT200 based GPUs are known to have a double 
precision flops peak 1/8th that of the single precision peak.  In practice, it should be 
noted that the GF100 cards run faster than 2x that of GT200 GPUs due to other architectural 
optimizations between the cards.  In particular, small kernels which have not been or do 
not parallelize as well run much faster on the GF100 GPUs.  
Lastly, in addition to our code supporting more devices and running 2x the speed, the code would be 
easier to optimize to ensure global memory coalescing and to prevent shared memory bank 
conflicts due to the 4 byte size of single precision floating point.  

However, software implementation of double precision has downsides.  Notably, it is often difficult 
to achieve the same precision as double precision floating point when using two single precision 
floating point values.  Additionally, full conversion 
of the code to use single precision floating point values and redefining basic vector operations would involve extensive reworking of 
the kernels that could lead to dozens of extra operations in each particular kernel.  It is not obvious 
that this sort of optimization is entirely suited to the GPU because increasing the number of 
operations in a kernel can increase register usage - an effect we are trying to avoid for optimization 
reasons.  Finally, simply modifying the distance calculation to be in double precision and 
computing the rest of the acceleration step in single precision would result in the largest 
performance benefit with the least amount of coding required but could potentially lose a large amount 
of precision in our calculations.

Due to the previously mentioned register restrictions and the complexity of our code, it is often 
difficult to achieve high occupancy on the devices we use.  It may be possible to rewrite entire 
routines in non-obvious ways to reduce the number of maximum registers in use at a given point in 
time to remedy this problem.  This is only a "medium priority" optimization and is probably the 
most difficult due to the number of ways kernels can be re-written.  This optimization is also 
hardware dependent and so may not be a worthwhile optimization.

\section{Summary and Conclusion}

We have described an implementation of a hybrid second order symplectic integrator for planetary 
system integration that permits close approaches.  It is similar in design to {\it SyMBA} \citep{duncan98} 
and {\it Mercury} \citep{chambers99} but is written in CUDA and works in parallel on a GPU.  The 
code is almost as accurate as the older integrators but is faster when many particles are
simultaneously integrated.  Bounded energy errors are observed during numerical integration of a 
few test cases implying that our integrator is indeed nearly symplectic.  The code has been
written primarily with parallel primitives so that modified versions can be written for other parallel 
computation platforms.  The current version of the code does not take into account collisions
between particles, however we plan to modify future versions so that these can be incorporated into 
the code.  We also plan to modify the encounter algorithms so that many dust particles can be 
integrated in the vicinity of planets and planetesimals.

\vskip 0.1 truein

Support for this work was provided by NSF through award AST-0907841.
We thank Richard Edgar for the design and initial set up of our GPU cluster.
We thank NVIDIA for the gift of four Quadro FX 5800 and two GeForce GTX 280 video cards.
We thank Nicholas Moore for informative discussions regarding GPU architectural 
and coding minutiae. 

\appendix
\section{Appendix: f and g functions}

Given a particle with position ${\bf x}_0$ and velocity ${\bf v}_0$ at time $t_0$
in Keplerian orbit,
its new position, $\bf x$, and velocity, $\bf v$, at time $t$ can be computed
\begin{eqnarray}
{\bf x} &=& f {\bf x}_0 + g{\bf v}_0 \nonumber \\
{\bf v} &=& \dot f {\bf x}_0 + \dot g{\bf v}_0
\end{eqnarray}
in terms of the $f$ and $g$ functions and their time derivatives, $\dot f$
and $\dot g$.
Introductory celestial mechanics textbooks often discuss $f$ and $g$ functions for particles solely in elliptic orbits.
However if a particle in a hyperbolic orbit is advanced using
elliptic coordinates, a NaN will be computed that can propagate via
the interaction steps.
It is desirable
to advance particle positions for all possible orbits, including parabolic or
hyperbolic ones.
This can be done by computing the $f$ and $g$ 
functions with universal variables, as described by \citet{prussing93}.
The recipe is repeated here as it is useful, but not available in  
most textbooks.

With $\mu \equiv \sqrt{GM}$ and $\alpha \equiv 1/a$ where $a$ is the semi-major
axis and $r_0$ the initial radius,
the $f$ and $g$ functions and their time derivatives in universal variables 
are computed as
\begin{eqnarray}
f &=& 1 - {x^2 \over r_0} C(\alpha x^2) \nonumber \\
g &=& (t - t_0)  - {x^3 \over \sqrt{\mu}} S(\alpha x^2) \nonumber \\
\dot f &=& {x \sqrt{\mu} \over r r_0} \left[\alpha x^3 S(\alpha x^2) -1 \right] \nonumber \\
\dot g &=& 1 - {x^2 \over r} C(\alpha x^2)
\end{eqnarray}
(see equations 2.38 by \citealt{prussing93}).
Here the variable $x$ solves the universal differential Kepler equation
\begin{equation}
\sqrt{\mu} (t-t_0) = {({\bf r}_0 \cdot {\bf v}_0) x^2 \over \sqrt{\mu}}
C(\alpha x^2) + (1 - r_0 \alpha) x^3 S(\alpha x^2) + r_0 x,
\label{eqn:ukp}
\end{equation}
(equation 2.39 \citealt{prussing93}).
The $f$ and $g$ functions must be computed before the time derivatives, 
$\dot f$ and $\dot f$, so that 
$r$, the radius at time $t$, can be computed.  This radius is then used
to compute $\dot f$ and $\dot g$.

The needed transcendental functions are
\begin{equation}
C(y) = 
\begin{cases}
        {1 \over 2!} - {y \over 4!} + {y^2 \over 6!} - \dots 
                & \mbox{if} ~~ y \sim 0  \nonumber \\
        {1 - \cos \sqrt{y} \over y }  
                & \mbox{if} ~~ y > 0  \nonumber \\
        {\cosh \sqrt{-y} - 1 \over -y }  
                & \mbox{if} ~~ y < 0  \\
\end{cases}
\end{equation}
and
\begin{equation}
S(y) = 
\begin{cases}
        {1 \over 3!} - {y \over 5!} + {y^2 \over 7!} - \dots 
                & \mbox{if} ~~ y \sim 0  \\
        {\sqrt{y} - \sin \sqrt{y} \over \sqrt{y^3} }  
                & \mbox{if} ~~ y > 0  \\
        {\sinh \sqrt{-y} - \sqrt{-y} \over \sqrt{-y^3} }  
                & \mbox{if} ~~ y < 0  \\
\end{cases}
\end{equation}
(equations 2.40 by \citealt{prussing93}).

It is convenient to define the function $F(x)$ and compute its derivatives
\begin{eqnarray}
F(x) &=& 
   -\sqrt{\mu} (t-t_0) + 
   {({\bf r}_0 \cdot {\bf v}_0) x^2 \over \sqrt{\mu}} C(\alpha x^2) 
   + (1 - r_0 \alpha) x^3 S(\alpha x^2) 
   + r_0 x \nonumber \\
F'(x) &= &
   {({\bf r}_0 \cdot {\bf v}_0) x \over \sqrt{\mu}} \left[1- \alpha x^2 S(\alpha x^2)  \right]
   + (1 - r_0 \alpha) x^2 C(\alpha x^2) 
   + r_0 \nonumber \\
F''(x) &= &
   {({\bf r}_0 \cdot {\bf v}_0)  \over \sqrt{\mu}} \left[1- \alpha x^2 C(\alpha x^2)  \right]
   + (1 - r_0 \alpha) x \left[ 1 - \alpha x^2 S(\alpha x^2)\right]  
\end{eqnarray}
(equation 2.41, 2.42 and problem 2.17 by \citealt{prussing93}).

To solve the universal Kepler equation (equation \ref{eqn:ukp}, that can
now be written as $F(x)=0$)
the Laguerre algorithm can be computed
iteratively \citep{conway} as
\begin{equation}
x_{i+1} = {n F(x_i) 
           \over 
          F'(x_i) \pm \left|(n-1)^2 F'(x_i)^2 - n(n-1) F(x_i) F''(x_i) \right|^{1/2}
          }
\end{equation}
(equation 2.43 \citealt{prussing93}).
Good numerical performance is found with $n=5$ \citep{conway}.
The sign in the denominator is the same as the sign of $F'(x_i)$.
The rate of convergence is cubic and convergence is achieved for any
starting value of $x$ \citep{conway}.  
Using a starting value $x_0=r_0$ we achieved convergence for a wide range
of timesteps and orbital parameters to a level of $10^{-16}$ 
in under 6 iterations.  Improved choices for starting values 
of $x$ are discussed by \citet{prussing93} and \citet{conway} but
require more computation than $x_0=r_0$.

{}

\end{document}